\documentstyle[sprocl,epsfig,rotate]{article}

\def\be{\begin{equation}}
\def\ee{\end{equation}}
\def\bea{\begin{eqnarray}}
\def\eea{\end{eqnarray}}

\newcommand{\nc}{\newcommand}

\nc{\bib}{\bibitem}
\nc{\1}{16^3\times{32}}
\nc{\2}{24^3\times{36}}
\nc{\napefv}{n_{\rm{ape}}(0.55)}
\nc{\nausfv}{n_{\rm{aus}}(0.55)}
\nc{\napeprm}{n_{\rm{ape}}(\alpha')}
\nc{\nausprm}{n_{\rm{aus}}(\alpha')}
\nc{\ncl}{n_c}
\nc{\nape}{n_{\rm{ape}}(\alpha)}
\nc{\naus}{n_{\rm{aus}}(\alpha)}
\nc{\napep}{n_{\rm{ape}}(\alpha')}
\nc{\nausp}{n_{\rm{aus}}(\alpha')}
\nc{\sutwo}{SU_{c}(2)}
\nc{\suthree}{SU_{c}(3)}
\nc{\sun}{SU_{c}(N)}
\nc{\szero}{S_{0}}
\nc{\szr}{\szero=8\pi^2/g^2}
\nc{\ql}{Q_{\rm{L}}(x)}

\begin{document}

\def\preprint{ADP-00-22/T405}
\def\archive{hep-lat/0004025}

\title{Visualizations of the QCD Vacuum}

\author{Derek B. Leinweber\footnote{Talk presented during the Workshop
on Light-Cone QCD and Nonperturbative Hadron Physics held at the CSSM,
U. Adelaide, Australia,  December 13--22, 1999.}}

\address{Special Research Center for the Subatomic Structure of
Matter (CSSM) and\\ Department of Physics and Mathematical Physics,
University of Adelaide 5005\\
E-mail: dleinweb@physics.adelaide.edu.au} 


\maketitle\abstracts{ Action and topological charge densities of the
Euclidean-space QCD vacuum are visualized in three-dimensional slices
of a $24^3 \times 36$ space-time lattice and an ${\cal
O}(a^2)$-improved $16^3 \times 32$ lattice.  Features include
instanton anti-instanton annihilation and a comparison of standard and
over-improved actions used in the smoothing of the gauge fields.  }

\section{Overview}\label{sec:overview}

The numerical approach to resolving the features of QCD is
occasionally criticized via statements\footnote{Such statements
overlook the fact that one can map out the properties of QCD by
exploring the quark-mass and temperature dependence of observables and
by resolving the individual quark sector contributions to various
observables.}  suggesting the knowledge of QCD lies ``in the
silicon.''  What these speakers are recognizing is that there are
massive amounts of data processed by today's supercomputers in
arriving at the final few bytes of information reported as QCD
observables.  The focus of this investigation is to further probe the
features of QCD by using visualization techniques which efficiently
convey the content of these massive amounts of data.

Here we will examine the properties of typical pure-gauge QCD-vacuum
field configurations created on a $24^3 \times 36$ space-time lattice
using the standard Wilson action at $\beta = 6.0$, which provides a
lattice spacing of 0.1 fm.  In addition, we consider an ${\cal
O}(a^2)$-improved action on a $16^3 \times 32$ lattice at $\beta =
4.38$ providing a lattice spacing of 0.17 fm.  To remove the
short-range noisy perturbative fluctuations, the field configurations
are smoothed by a local algorithm designed to minimize the gauge action at
each update.  This algorithm known as cooling \cite{cooling} is a well
established method for locally suppressing quantum fluctuations.  The
locality of the method allows topologically nontrivial field
configurations to survive numerous iterations of the cooling
algorithm.

Visualization techniques may be used to answer questions such as the
following: To what extent do the QCD vacuum field configurations
resemble a randomly oriented multi-(anti)instanton field
configuration.  Do the instantons display a remnant of spherical
symmetry expected for isolated instantons, or do the nonperturbative
interactions completely distort this picture?  While it has been
established that anti-instanton instanton pairs have an attractive
interaction and will annihilate during cooling, it is not clear how
this process takes place.  Do the pairs simply come together or do
they wrap around each other in the process of annihilation?  One can
also search for visual evidence of a polarization phenomena
\cite{Smith:1998wt} where large sized instantons tend to have on
average the same sign and are over screened by smaller instantons
which tend to have the opposite topological charge of the larger
instantons.  Visualization techniques have also been used to examine
the microscopic effects of various smoothing
algorithms.\cite{Bonnet:2000dc}

\section{Visualizations}

\begin{figure}[p]
\begin{center}
\epsfig{file=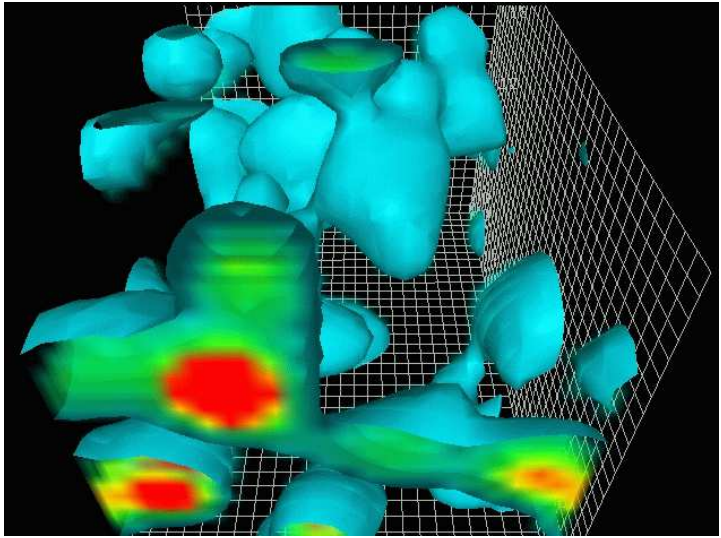,width=9.5cm}
\end{center}
\caption{The action density after 30 cooling sweeps.}
\label{WrappingAction}
\end{figure}

Fig.\ \ref{WrappingAction} illustrates a three-dimensional slice of
the QCD action density from the $24^3 \times 36$ lattice after 30
cooling sweeps.\footnote{Every link on the lattice is updated once
during a single sweep.}  Here the blue isosurface connects all points
having the same action density.  Tri-linear interpolation is used to
smooth the surface.  Volume rendering is done within the surface to
illustrate changes in the action density.  Outside the surface at low
action densities, no volume rendering is done in order to allow one to
see within the field configuration.  Sharp peaks in the action density
have been clipped to aid in the illustration.

\begin{figure}[p]
\begin{center}
\epsfig{file=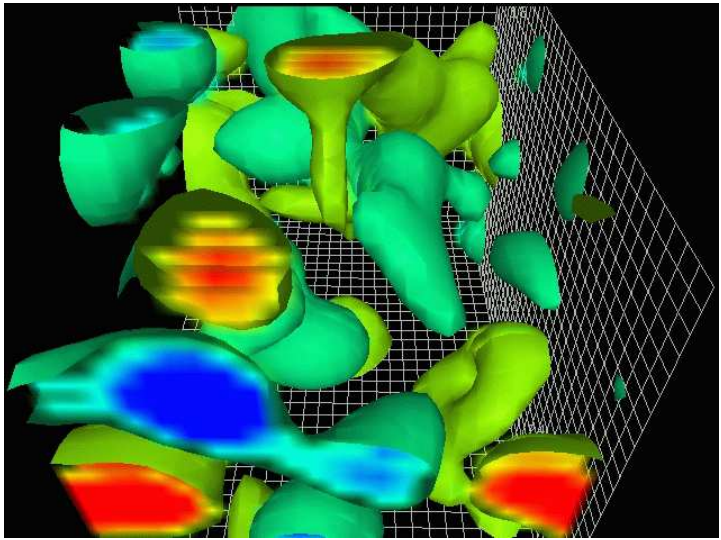,width=9.5cm}
\end{center}
\caption{The topological charge density after 30 cooling sweeps.}
\label{WrappingCharge}
\end{figure}

The regions in which the action density has survived have non-trivial
topological charge density as illustrated in Fig.\
\ref{WrappingCharge}.   Here the yellow (green) isosurface connects
points having equal positive (negative) topological charge density.
Volume rendering of the topological charge density inside the isosurfaces
illustrates the changes in the positive (negative) topological charge
densities by red (blue) extremes in colour.

Here we see that the (anti)instantons that are somewhat isolated do
show a remnant of spherical symmetry in the action and topological
charge densities.  Indeed similar plots in which the isosurface
explores very high action or topological charge density show an
elliptically shaped isosurface.

However, closely spaced (anti)instantons can show extreme deviations
from spherical symmetry.  The action and topological charge densities
illustrated in the upper center of Figs.\ \ref{WrappingAction} and
\ref{WrappingCharge} correspond to instanton anti-instanton pairs in
the process of annihilation.  By 100 iterations of the cooling
algorithm, the action and topological charge densities have largely
vanished from this region.  In the process of annihilation, one can
see regions in which fingers of significant topological charge density
are wrapping around negative topological charge density.  Between
these isosurfaces, the topological charge density is rapidly going to
zero and therefore is not rendered.  The action density simply shows a
region of significant interaction.

Animations (in animated-gif format suitable for viewing within a
browser) illustrating this process of
annihilation are available on the web at

\noindent
{\small\sf
http://www.physics.adelaide.edu.au/theory/staff/leinweber/VisualQCD/QCDvacuum/}\hfill\break
The process of cooling smoothes and extends the size of
(anti)instantons such that the action and charge densities decrease.
If one produces an animation with a fixed isosurface value one will
see the rendered regions shrink in size and eventually disappear as
the local density drops below the isosurface threshold.  One would
incorrectly reach the conclusion that instantons shrink during cooling
and eventually fall through the lattice.  These animations illustrate
the local action and charge densities relative to the total action and
topological charge to more correctly illustrate the distribution of
action and charge on the lattice.

\section{Over-Improved Cooling}

It is well known that errors in the standard Wilson action eventually
destroy (anti)instanton configurations.  Errors in the standard Wilson
action underestimate the action of a field configuration such that
application of a cooling sweep on a single instanton configuration
will result in an action less than the one-instanton bound.  The
instanton is spoiled by the cooling process and eventually the
remaining action will be eliminated via the cooling procedure.

Hence we also examine the five-loop over-improved action of De
Forcrand {\it et al.}~\cite{deforcrand} designed to render instantons
stable over several hundreds of sweeps.  This approach includes
extended planar paths combined to reproduce the classical action with
no ${\cal O}(a^2)$ nor ${\cal O}(a^4)$ discretization errors and with
coefficients fine-tuned to stabilize instantons over many applications
of the algorithm.

\begin{figure}[p]
\begin{center}
\epsfig{file=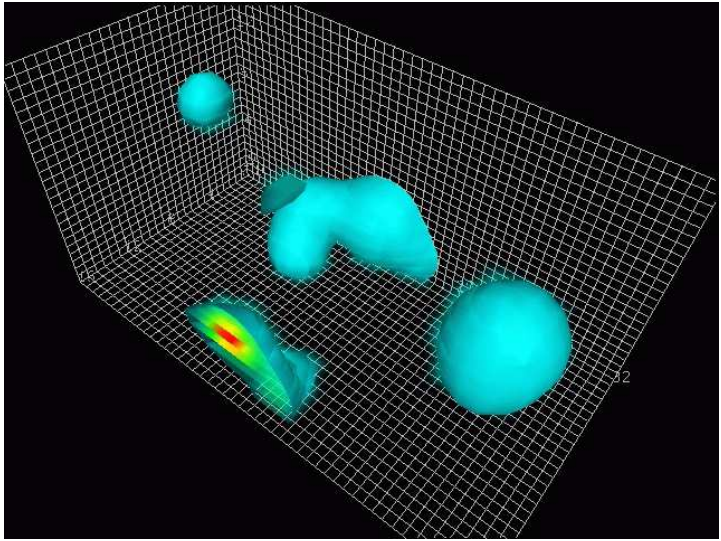,width=9.5cm}
\end{center}
\caption{The action density after 30 cooling sweeps using the standard
one-loop Wilson action.}
\label{OneLoopAction}
\end{figure}

To investigate these algorithms based on improved actions, we consider
pure-gauge configurations from an ${\cal O}(a^2)$-improved $16^3
\times 32$ lattice with a lattice spacing of 0.17 fm.  Fig.\
\ref{OneLoopAction} illustrates a three-dimensional slice of the
action density after 30 sweeps of cooling using the standard one-loop
Wilson action.  The population of (anti)instantons is very sparse and
the isolated instantons are quite spherical.

\begin{figure}[p]
\begin{center}
\epsfig{file=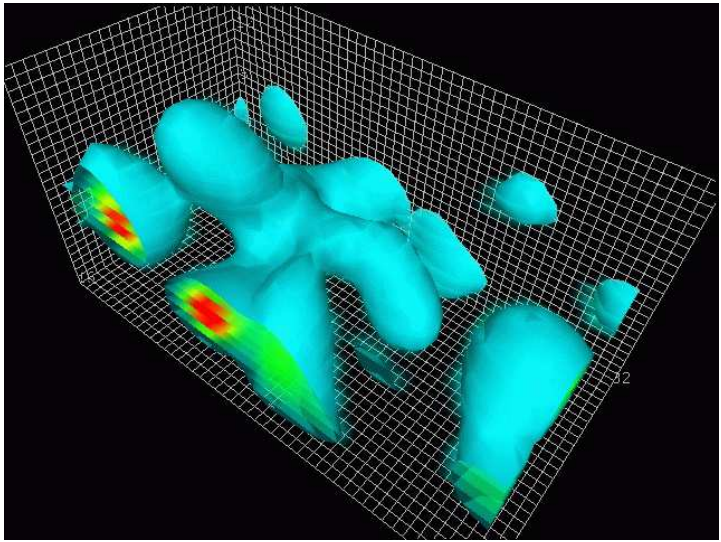,width=9.5cm}
\end{center}
\caption{The action density after 30 cooling sweeps using the
five-loop over-improved action.\protect\cite{deforcrand}}
\label{FiveLoopAction}
\end{figure}

Fig.\ \ref{FiveLoopAction} illustrates the same gauge-field
configuration, this time cooled with a parallel implementation
\cite{Bonnet:2000db,sundance} of the five-loop over-improved action.
While there is a correspondence between the instantons surviving 30
sweeps of the one-loop action and those of the five-loop action, it is
clear that many more (anti)instantons have survived the improved
cooling algorithm.

Hence one can {\it see} the difficulty in cooling with the standard
one-loop Wilson action.  The density and size of (anti)instantons is
dependent on the number of cooling sweeps.\footnote{While these facts
are fairly well known, we find the visualizations illustrating these
difficulties very compelling.}  The true vacuum is much denser as
suggested by the five-loop action density.

In short, the visualizations reveal rich structure in typical field
configurations of the QCD vacuum.  The action and topological charge
densities presented here display long-range non-perturbative
correlations between strongly interacting instantons and
anti-instantons. 

\section*{Acknowledgments}

Thanks to John Ahern, Sundance Bilson-Thompson, Frederic Bonnet,
Patrick Fitzhenry, Greg Kilcup, Mark Stanford, Tony Williams and James
Zanotti for their contributions to making these visualizations
possible.  Additional thanks to Francis Vaughan of the SACPC for
generous supercomputer support and the DHPC Group for support in the
development of parallel algorithms.  This research is supported by the
Australian Research Council.

\end{document}